\documentclass[twocolumn,prl,aps,epsf,floats]{revtex4}

    \newif \ifpdf
    \ifx \pdfoutput\undefined
    \pdffalse 
    \else
    \pdfoutput=1 
    \pdftrue
    \fi

    \ifpdf
    \usepackage[pdftex]{graphicx}
    \else
    \usepackage{graphicx}
    \fi

\begin{document}

 \ifpdf
    \DeclareGraphicsExtensions{.pdf, .jpg, .tif}
    \else
    \DeclareGraphicsExtensions{.eps, .jpg}
    \fi


\title{Evidence of a first order phase transition between Wigner crystal and Bubble phases of 2D electrons in higher Landau levels}
\author{R. M. Lewis$^{1,2}$, Yong Chen$^{1,2}$, L. W. Engel$^1$, D. C. Tsui$^2$, P. D. Ye$^{1,2}$, L. N. Pfeiffer$^{3}$, and K. W. West$^{3}$
}

\address{$^{1}$NHMFL, Florida State University, Tallahassee, FL 32310, USA\\
$^{2}$Dept.\ of Electrical Engineering, Princeton University, Princeton, NJ 08544\\
$^{3}$Bell Laboratories, Lucent Technologies, Murray Hill, NJ 07974}

\date{\today}
\begin{abstract}
For filling factors $\nu$ in the range between 4.16 and 4.28, we simultaneously detect {\it two} resonances in the real diagonal microwave conductivity of a two--dimensional electron system (2DES) at low temperature $T \approx 35$ mK. We attribute the resonances to Wigner crystal and Bubble phases of the 2DES in higher Landau Levels. For $\nu$ below and above this range, only single resonances are observed.  The coexistence of both phases is taken as evidence of a first order phase transition.  We estimate the transition point as $\nu=4.22$.

\end{abstract}

\maketitle    
\vskip1pc

    Clean two dimensional electron systems (2DES) show strong evidence of several solid phases in moderate magnetic fields ($B$), when the filling factor $\nu= n h /eB > 4$, and the temperature $T <100$ mK.  These phases are characterized by the partial filling of the uppermost spin split Landau level (LL), $\nu^* = \nu-[\nu]$ where $[\nu]$ is the greatest integer less than $\nu$.  The density of electrons in the uppermost LL is given by $ n^*= n \nu^*/\nu$.
When $\nu^* $ is small, the inter--electron spacing is large compared with the cyclotron radius ($1/ \sqrt{ n^*} >> r_c$) and a Wigner crystal phase \cite{wc} with triangular lattice is expected to form the ground state\cite{macgirvin,fogler,yoshioka,kunyang}.  This crystalline phase would be pinned by residual disorder and therefore would be insulating.  A microwave resonance \cite{ychen,rlewis2} was detected in the frequency dependence of the real diagonal  conductivity (Re$[\sigma_{xx}(f)]$) around integer $\nu=1,2,3,$ and 4 for $\nu^* < 0.2$. This resonance has been interpreted as the pinning mode of such an integer quantum Hall Wigner crystal (IQHWC).

	For $\nu^* $ above $\sim 0.2$, a new crystal phase with clusters of two or more electrons per lattice site---dubbed the bubble phase (BP)---is predicted\cite{fogler,kunyang,yoshioka}.  The BP is favored energetically once the single particle wavefunctions show significant overlap,  $1/  \sqrt{ n^*} \sim 2 r_c$ \cite{fogler}.   DC transport experiments \cite{mlilly,rrdu,cooper_iv} find minima in the diagonal resistance ($R_{xx}$) together with Hall resistance quantized to the value of the adjacent integer plateaus at $\nu \sim 4 +1/4$ and $4+3/4$.   Non--linear current--voltage characteristics \cite{cooper_iv}, a microwave resonance  \cite{rlewis}, and narrow band noise above a threshold voltage\cite{cooper_noise} have been seen around these $\nu$.  Collectively, these measurements are strong evidence for the BP.  Ultimately, near $\nu^* \sim 1/2$, the 2DES is expected to form a stripe state \cite{fogler}.  Anisotropic transport measurements \cite{mlilly,rrdu} have provided strong support for stripes.  This sequence of phases is repeated in reverse for $\nu^* >1/2$.

    Microwave conductivity measurements \cite{ychen,rlewis2,rlewis}, like the ones we present here, detect a sharp resonance in Re$[\sigma_{xx}(f)]$ when the 2DES is in a solid phase.  A similar microwave resonance has been known for some time to occur in  2DES in the insulating phase \cite{lloydssc,peidewc} at $\nu < 1/5$ and low $T$ where theory favors the formation of a Wigner Crystal \cite{wc}.  Such resonances are interpreted as due to the pinning mode\cite{fukuyama_lee} of the electron solid---the oscillation of crystalline domains, pinned by disorder, in the zero wavevector limit. 

    In this paper, we present measurements of Re$[\sigma_{xx}(f)]$ which focus on the transition between the the IQHWC and the BP regimes.  Theory has suggested that this transition is first order \cite{yoshioka, fogler_review, goerbig, cote}. In the range of $\nu$ between 4.05 and 4.28  and $f$ between $\sim$ 2 GHz and 700 MHz, we detect a resonance in Re$[\sigma_{xx}(f)]$ which we attribute to the IQHWC \cite{ychen, rlewis2}. For $\nu$ between 4.16 and 4.41, and at $f$ between 500 MHz and 150 MHz, we detect a second resonance in Re$[\sigma_{xx}(f)]$ which we attribute to the BP \cite{rlewis,cooper_iv,cooper_noise}.  For a narrow range of $\nu$, both resonances are {\it simultaneously} present, implying the coexistence of both phases.  This coexistence signals a first order phase transition between the IQHWC and BP as $\nu^*$ moves from 0.16 to 0.28. 

The sample used for these measurements is a 300\AA\  GaAs/AlGaAs quantum well grown by molecular beam epitaxy.  It has density $n=3.0 \times 10^{11}\  {\rm cm^{-2}}$ and mobility $\mu=2.4 \times 10^7 \ {\rm cm^2 \  V^{-1} \ s^{-1}}$ at 300 mK.  The 2DES is approximately 2000\AA \ below the surface.   A coplanar wave guide (CPW) \cite{wen} is patterned onto the sample surface using standard photolithography and thermal evaporation.  A CPW consists of a driven center line separated from grounded planes on either side by slots of width $w$.  In the absence of the 2DES, the CPW on the sample would have characteristic impedance ($Z_0$) of 50 $\Omega$.   At $B$ fields large enough for the quantum Hall Effect to manifest itself, the small conductivity of the 2DES appears as a weak loading of the CPW and some of the microwave power traversing the CPW is absorbed \cite{lloydprl93}.  In this low loss limit, the real part of the diagonal conductivity is given by Re$[\sigma_{xx}]= -\frac{w}{2lZ_0} \ln (P_t/P_0)$  where $P_t$ is the transmitted power, $P_0$ is the power transmitted in the absence of the 2DES, and $l$ is the length of the CPW.  The data presented here were measured using a CPW pattern  where $l=28$ mm and $w=30 \mu$m.  All the data were measured at $T\approx 35$ mK in the low power limit, in which decreasing rf power does not alter data except to decrease signal.

\begin{figure}[tb]
\begin{center}
\includegraphics[width=8.5cm]{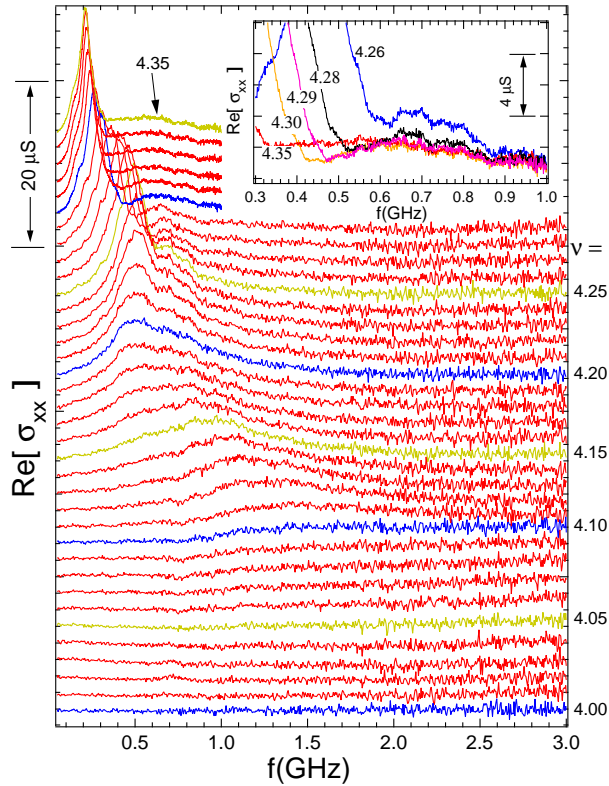}
\caption{The real part of the diagonal conductivity, Re[$\sigma_{xx}$] vs frequency, $f$, at $\nu$ between 4.00 and 4.35 in steps of 0.01. Traces are offset in increments of 2 $\mu$S. $\nu$ is marked on the right axis, and the temperature, $T \approx 35$ mK. {\bf Inset} are  Re$[\sigma_{xx}]$ vs $f$  at $\nu=4.26$, 4.28, 4.29, 4.30, and 4.35 with the vertical scale expanded.}
\label{fig.1}
\end{center}
\end{figure}

In Fig.\ 1 we show 36 traces of  Re[$\sigma_{xx}$] versus $f$, the first 30 for $\nu$ between 4.00 and 4.29 ($f$ between 50 MHz and 3 GHz), and a subsequent six between $\nu= 4.30$ and 4.35 ($f$ between 50 MHz and 1 GHz).  The traces are offset in increments of 2 $\mu$S.  Close to $\nu=4.00$, Re$[\sigma_{xx}]$ is nearly featureless.  At $\nu=4.10$, a rise in Re$[\sigma_{xx}]$ near 1.5 GHz is evident, and at $\nu=4.11$, Re$[\sigma_{xx}]$ shows a weak resonance with a peak at $f_{p1} \approx1.46$ GHz.  The position of this resonance shifts to lower frequency in the next few traces.  By $\nu=4.15$, $f_{p1} \approx 900$ MHz and the the full width at half maximum, $\Delta f \approx 1$ GHz, can be estimated by fitting a lorentzian curve.  However, as we continue to move to higher $\nu$, the shape of the resonance distorts.  At $\nu=4.18$ it is almost square and at 4.19 it takes on a double bump.  Starting at about $\nu=4.20$, a sharp new peak is visible at $f_{p2} \approx 470$ MHz.  From $\nu=4.20$ to 4.25, $f_{p2}$ remains roughly constant (between 470 and 500 MHz) in the traces as does $f_{p1} \approx 700$ MHz.  From $\nu=4.26$, $f_{p2}$ changes rapidly as $\nu$ is increased. $f_{p2}$ is 450 MHz at $\nu=4.26$, has decreased to 320 MHz by $\nu=4.29$, and is only about 210 MHz at $\nu=4.35$.  The error in determining $\nu$ on the data in Fig.\ 1 is $\pm 0.01$ and is primarily caused by flux trapped in the superconducting magnet. 

Inset into Fig.\ 1, we show traces of  Re$[\sigma_{xx}]$ versus $f$ at $\nu=4.26$, 4.28, 4.29, 4.30, and 4.35 with an expanded vertical scale to illustrate how the lower $\nu$ resonance, at $f_{p1}$, vanishes.  At $\nu=4.26$ the data show a peak at $f_{p1} \approx 690$ MHz.  The upturn below 550 MHz is due to the second resonance at $f_{p2}$.  At $\nu=4.28$, Re$[\sigma_{xx}]$ is greater at $f$ near 650 MHz than at $\nu=4.29$.  However, Re$[\sigma_{xx}]$ for $f$ from 600 MHz to 1 GHz is the same for $\nu=4.29$, 4.30, and 4.35.  Based on this criterion of unchanging Re$[\sigma_{xx}]$, the highest $\nu$ at which we detect $f_{p1}$ is $\nu=4.28$.
 
The two resonances in Re[$\sigma_{xx}$] seen in Fig.\ 1 have been identified previously.  The lower frequency peak, $f _{p2} $, was interpreted by Lewis {\it et al.} \cite{rlewis} as  the pinning mode of the BP which manifests itself  between $\nu \approx 4.2$ and 4.4 for $T< 100$ mK.  In what follows, we refer to this as the BP peak.  The higher frequency peak, $f_{p1}$, occurs between $f=700$ MHz and $\sim 2$ GHz and is present between $\nu \sim 4.05$ and 4.28.  It is understood as the pinning mode of the IQHWC\cite{ychen}.

\begin{figure}[tb!]
\begin{center}
\includegraphics[width=8cm]{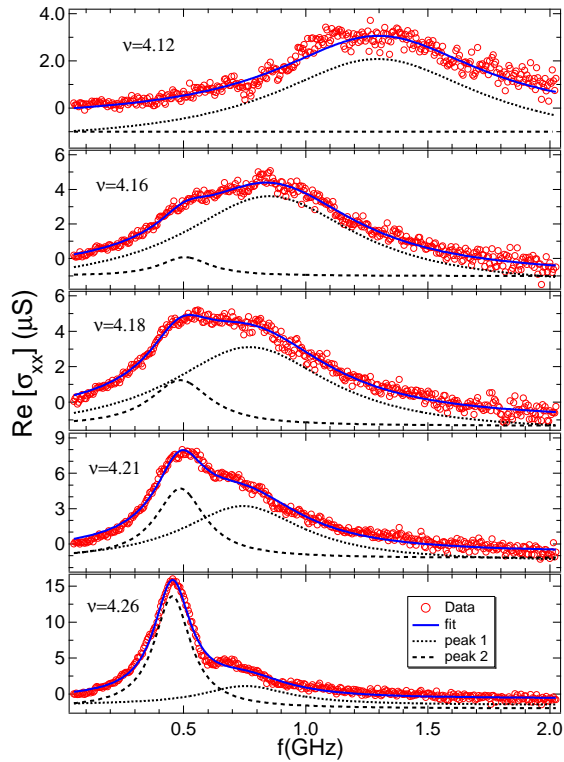}
\caption{  Re[$\sigma_{xx}$] versus $f$ from 50 MHz to 2.025 GHz at $\nu=4.12,\ 4.16, \  4.21,\ {\rm and} \ 4.26$. Two peak lorentzian fits to this data are shown as the total fit (solid line over the data) and the individual peaks (dashed lines offset 1 $\mu$S below).}
\label{fig.2}
\end{center}
\end{figure}

In Fig.\ 2 we plot traces at $\nu=4.12$, 4.16, 4.18, 4.21, and 4.26 together with a two peak lorentzian \cite{lorentzian} fit for $f$ between 50 MHz and 2.025 GHz.  The data are shown as open circles, the individual peaks are plotted as dashed lines offset $1 \mu$S below the data, and the total fit is shown as a solid line.  The fits are made to the data shown in Fig.\ 1 after subtracting a non--resonant background $g(f)$ given by $g(f)=1\mu {\rm S/GHz} \times f$ for $f < 2$ GHz and $g(f)= 2 \mu$S (constant) for $f> 2$ GHz.  At $\nu=4.12$, the data are well fit by a single lorentzian with $f_{p1} = 1.3 \pm 0.07$ GHz.  The full two--peak lorentzian fit is necessary at $\nu=4.16$ where the BP peak contributes at $f_{p2}=505$ MHz, $\Delta f_{p2} \approx 250$ MHz, and peak conductivity ($\sigma_{p2}$) of 1.1 $\mu$S.  At the same $\nu$, $f_{p1} = 852 \pm 50$ MHz, $\Delta f_{p1} = 870$ MHz.  For $\nu=4.18$, the fit finds $ f_{p2}= 485 \pm 20$ MHz, and $\sigma_{p2} \approx 2.6 \ \mu$S and easily separates $f_{p1}$ at $808 \pm 80$ MHz.  At $\nu=4.21$, the BP peak is larger than the IQHWC peak and the fit finds $f_{p2}= 486 \pm 20$ MHz, $\sigma_{p2} \approx 6 \ \mu$S, $f_{p1}= 741 \pm 40$ MHz, and $\sigma_{p1} \approx 4.5 \ \mu$S.  At $\nu=4.26$, the BP peak has moved to $f_{p2} = 454$ MHz leaving the IQHWC peak at $ f_{p1} = 758 \pm 30$ MHz. The lorentzian curves describe the data well, and are adequate for extracting the position of each peak, approximate line widths, and the weight or integrated strength of each peak.

\begin{figure}[tb!]
\begin{center}
\includegraphics[width=9cm]{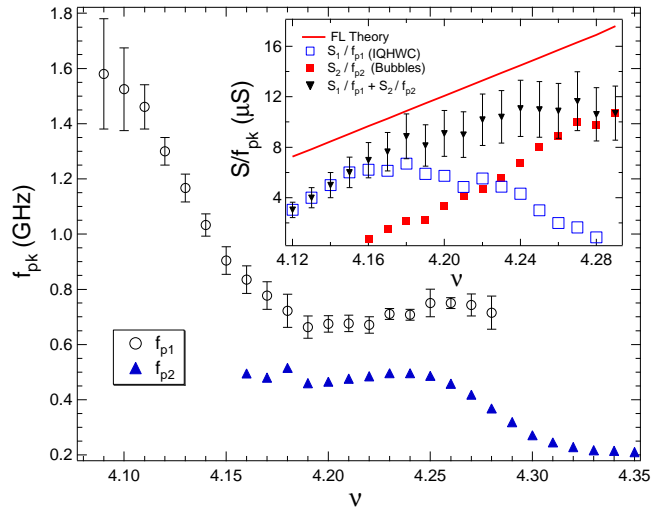}
\caption{Peak frequency for both resonances, $f_{pk}$,  versus filling factor, $\nu$, from 4.08 to 4.35. {\bf Inset} is the oscillator strength ($S$) divided by $f_{pk}$ for each resonance vs. $\nu$ as determined from the fits.  The solid line is from theory by Fukuyama and Lee\cite{fukuyama_lee}.  The other symbols are the individual peaks and their sum.}
\label{fig.3}
\end{center}
\end{figure}

In Fig.\ 3 we plot $f_{p1}$ and $f_{p2}$, obtained through fitting the data in the manner shown in Fig.\ 2 \cite{fitsnote}, against $\nu$.  Three distinct regions are seen, which we characterize by $\nu$.  In the first, from $\nu \approx $ 4.05 to 4.16, only the IQHWC resonance appears.  Here, $f_{p1}$ lies between 2 GHz and 700 MHz and decreases rapidly as $\nu$ increases.  The second range occurs between $\nu=4.16$ and 4.28 where the two resonances are both clearly present.
  Within the second region is a subrange of $\nu$, between 4.19 and 4.25, where $f_{p1}$ and $f_{p2}$ are both nearly constant.  In this subrange, $f_{p1} \approx 730$ MHz and $f_{p2} \approx 500$ MHz.  Finally, in the third region, for $\nu > 4.28$, the BP resonance dominates the spectrum, $f_{p2}$ decreases rapidly from about 370 MHz at $\nu=4.28$ down to 210 MHz at $\nu=4.35$. The BP resonance is present to $\nu \approx 4.41$.  The errors on $f_{p1}$ are shown and those on $f_{p2}$ are less than 20 MHz, about the same size as the symbols.  The total error in determining both peaks is less than the difference ( $f_{p1}-f_{p2}$) in all cases.

Inset into Fig.\ 3 is $S/f_{pk}$ versus $\nu$, where $S$ is the integral of Re$[\sigma_{xx}(f)]$.  For a classical Wigner crystal of density $n_{wc}$, Fukuyama and Lee (FL) \cite{fukuyama_lee} predict $S/f_{pk}= n_{wc} e \pi / 2 B$. Substituting $ n^*$ for $n_{wc}$ gives the solid line in the figure.
 Integrating the fit to the IQHWC peak yields $S_1/f_{p1}$ which has a maximum  at $\nu \approx 4.17$.  $S_1/f_{p1} < 2 \mu$S above $\nu=4.25$, although some weight remains in the IQHWC even at $\nu=4.28$.  $S_2/f_{p2}$ is the analogous quantity calculated from the fit to the BP peak. $S_2/f_{p2}$ is 0 below $\nu=4.16$, but  increases smoothly thereafter to the highest $\nu$ shown.  The two curves cross at $\nu=4.22$.  The sum of IQHWC and BP contributions, $S_1/f_{p1} + S_2/f_{p2}$, indicates that $n^*$ is approximately the density of electrons participating in the IQHWC and BP \cite{ychen,rlewis}.  $S_1/f_{p1}+ S_2/f_{p2}$ is within 50 \% of the FL result\cite{fukuyama_lee} and has an error $\sim$20\%.

	The simultaneous occurrence of both resonances between $\nu=4.16$ and 4.28 is not due to macroscopic inhomogeneity of $n$ in the sample.  Primary evidence of homogeneity can be seen in the sharp BP resonance which at $\nu=4.25$ has $f_{p2}=210$ MHz, and $\Delta f \approx 75$ MHz.  Furthermore, between $\nu=4.26$ and 4.32, $f_{p2}$ is exquisitely sensitive to $B$ and yet always has a single value.  Similarly, the IQHWC resonance also shows strong dependence on $B$ between $\nu=4.09$ and 4.16.  Only in the range of $\nu$ between 4.16 and 4.28 are both resonances seen.  Finally, measurements of this sample detected fractional quantum Hall states comparable to those seen in previous measurements\cite{ychen,rlewis, rlewis2} of this wafer.

One signature of a first order phase transition is the coexistence of both phases at the transition point.  This is not seen in a second order phase transition where phases meet i.e., become identical, at the transition.  Our data are consistent with a transition from the IQHWC to the BP taking place between $\nu=4.16$ and 4.25.  Throughout this range, both resonances are seen in coexistence, which we take as strong evidence that a first order phase transition occurs.  The crossover in $S/f_{pk}$ occurs at $\nu=4.22 \pm 0.02$.  We are unable to find hysteresis associated with this transition in our measurements.

A first order transition between IQHWC and the BP has been explicitly predicted by Shibata and Yoshioka \cite{yoshioka}. The possibility of a first order phase transition between the IQHWC and the BP and between the BP and stripe phase with regions of coexistence about the transitions has been suggested by Fogler\cite{fogler_review}.   Other Hartree--Fock calculations show discontinuous jumps in the magnetic moment and magnetic susceptibility \cite{cote} at the transition between the IQHWC  and BP.  More recently, Goerbig {\it et al.} \cite{goerbig}, using a model incorporating a short range Gaussian disorder potential and a classical shear modulus, have calculated that $f_{pk}$ should drop in proportion to $\sqrt{2}$ in moving from the IQHWC to the two electron BP.  They also suggest coexistence of the two phases close to the transition may allow for the observation of two peaks in Re$[\sigma_{xx}]$.  The $\approx$ 30\% drop we observe in $f_{pk}$ is remarkably consistent with the IQHWC to BP transition in their model and suggests that the BP around $\nu=4 +1/4$ has two electrons per lattice site. 

The crossover in $S/f_{pk}$ from IQHWC to BP occurs around $\nu=4.22 $.  This is in excellent agreement with the Hartree--Fock calculations of Goerbig {\it et al.\ } \cite{goerbig} and Cote {\it et al.\ } \cite{cote} who predict the transition at $\nu \approx 4.21\ {\rm and}\ 4.22 $ respectively and with calculations by Shibata and Yoshioka \cite{yoshioka} whose  density matrix renormalization group method predicts a transition at $\nu=4.24$.

For $\nu^*$ well within either the IQHWC or BP regions, $f_{pk}$ decreases smoothly as $\nu^*$ increases.  This dependence of $f_{pk}$ upon $\nu$ is in agreement with results in Refs.\ \cite{ychen,rlewis} and consistent with weak pinning theories \cite{fukuyama_lee,chitra} for 2DES in high $B$.  In these theories,  $f_{pk}$ {\it decreases} as the domain size ($L$) or the shear modulus ($\mu$) {\it increases}, and both $L$ and $\mu$ increase as $n^*$ increases.  In this context, we may interpret the decrease in $f_{pk}$ as $\nu^*$ increases as due to the steady increase in $n^*$.

The IQHWC and BP peak frequencies are relatively constant in much of the coexistence range. Speculating that $f_{p1}$ and $f_{p2}$ vary due to the density of their respective phases, we propose the following model. 
The IQHWC reaches a maximum density ($n_1$) at $\nu \approx 4.19$.  Likewise, the BP has minimum density ($n_2$) at $\nu \approx 4.25$.  For $\nu$ in between, $n^*= \alpha(\nu) n_1 + (1-\alpha) n_2$ where $\alpha(\nu)$ is the percentage of $n^*$ participating in the IQHWC.   Thus, increasing $\nu$ from 4.19 to 4.25 results in a shift in $S$ from the IQHWC peak to the BP peak.  At finite $T$, domain walls in such an admixture might contribute to the finite dc resistance which is seen\cite{cooper_iv,ggervais} at $\nu \sim 4.20$.

We have shown evidence that the transition between the IQHWC and BP is of first order and that it occurs between $\nu=4.16$ and 4.28.  Between these $\nu$,  both resonances are simultaneously present in Re$[\sigma_{xx}(f)]$ indicating the coexistence of the IQHWC and BP.  For $\nu < 4.16$ down to about 4.05, only the IQHWC resonance is present. For $\nu > 4.25$ up to about 4.40, the BP resonance dominates.  From the intensities of the resonance, we estimate the crossover in $S/f_{pk}$ from IQHWC to BP occurs at $\nu=4.22 \pm .02$.  

We thank Kun Yang and Herb Fertig for stimulating discussions and Eric Shaner for help making a mask.  The measurements were performed at the NHMFL with support from the AFOSR and the NHMFL in--house research program. The NHMFL is supported by DMR--0084173 and by the State of Florida.


\begin{thebibliography}{9}
\bibitem{lorentzian} The functional form we use is: $g(f)= A_0 /((f_{pk}-f)^2 + \Gamma)$ where $A_0$ is the amplitude, $f_{pk}$ is the peak frequency and one finds $\Delta f = 2  \sqrt{\Gamma}$.
\bibitem{fitsnote}The subtraction of non-resonant backgrounds from the data does not, within error, change the $f_{pk}$ determined from the fits.
\end{thebibliography}
\end{document}